\documentclass{emulateapj}

\newcommand{\sbr}[1]{_{\mathrm{#1}}}

\begin{document}
\submitted{ApJ in press}

\author{R. W. Pike and Michael J. Hudson\altaffilmark{1}}
\title{Cosmological Parameters from the Comparison of the 2MASS
Gravity Field with Peculiar Velocity Surveys} \affil{Dept of Physics,
University of Waterloo, Waterloo, ON, N2L 3G1 Canada}
\altaffiltext{1}{email: mjhudson@uwaterloo.ca}

\begin{abstract}
We compare the peculiar velocity field within 65 $h^{-1}$ Mpc
predicted from 2MASS photometry and public redshift data to three
independent peculiar velocity surveys based on type Ia supernovae,
surface brightness fluctuations in ellipticals, and Tully-Fisher
distances to spirals. The three peculiar velocity samples are each in
good agreement with the predicted velocities and produce consistent
results for $\beta_{K}=\Omega\sbr{m}^{0.6}/b_{K}$. Taken together the
best fit $\beta_{K} = 0.49 \pm 0.04$.  We explore the effects of
morphology on the determination of $\beta$ by splitting the 2MASS
sample into E+S0 and S+Irr density fields and find both samples are
equally good tracers of the underlying dark matter distribution, but
that early-types are more clustered by a relative factor
$b\sbr{E}/b\sbr{S} \sim 1.6$. The density fluctuations of 2MASS
galaxies in $8\,h^{-1}$ Mpc spheres in the local volume is found to be
$\sigma\sbr{8,K} = 0.9$. From this result and our value of
$\beta_{K}$, we find $\sigma_8 (\Omega\sbr{m}/0.3)^{0.6} =
0.91\pm0.12$.  This is in excellent agreement with results from the
IRAS redshift surveys, as well as other cosmological probes. Combining
the 2MASS and IRAS peculiar velocity results yields $\sigma_8
(\Omega\sbr{m}/0.3)^{0.6} = 0.85\pm0.05$.
\end{abstract}

\section{Introduction}

Peculiar velocities are a unique probe of the distribution of mass in
the nearby universe. The velocity of an object, such as a galaxy, is
the sum of two contributions: the cosmological expansion and the
peculiar velocity, which arises from gravitational attractions from
surrounding overdensities, which are dominated by dark matter.

In the linear regime, the peculiar velocity ${\bf{v}}(\bf{r})$ is
given by
\begin{equation}
{\bf{v}}\left( {\bf{r}} \right) =
\frac{\Omega\sbr{m}^{0.6}}{4\pi }\int d^3 {\bf{r}}^{\prime
} \delta\sbr{m}\left( \mathbf{r}^{\prime }\right) \frac{\left( \mathbf{r}^{\prime
}-\mathbf{r}\right) }{\left| \mathbf{r}^{\prime }-\mathbf{r}\right| ^3}.
\label{peculiar}
\end{equation}
where $\delta\sbr{m} \left( \mathbf{r}\right) = ({\rho
  -\overline{\rho}})/{\overline{\rho}}$, and $\overline{\rho}$ is the
average density of the universe. Equation (\ref{peculiar}) is a
byproduct of the assumption that structure forms as a result of the
growth of small inhomogeneities in the initial density
field. Formally, it is valid only in the linear regime where $\delta
\lesssim 1$ i.e., on scales larger than $\sim 5$ Mpc.  With an all-sky
mass density field, the resulting velocity field can be predicted and
compared to observations.  Note that this velocity-velocity (v-v)
comparison method requires a complete description of the density
field, in order to use Equation (\ref{peculiar}).

There are two common approaches to estimate $\delta\sbr{m}$.  One approach
is to assume simple parametric infall models (e.g.\ ``Virgo infall''),
and fit for parameters of the model. The second approach, adopted
here, is to assume that galaxies are tracers of the mass density
field. In this context, it has become a common practice to employ the
simplifying assumption of linear biasing, in which $\delta\sbr{g}=b\,\delta
\sbr{m}$, where $\delta\sbr{g}$ is the galaxy density contrast and $b$ is the
bias factor relating the mass-tracer (galaxy) fluctuations with the
mass fluctuation field.  With the inclusion of linear biasing, there
is a relationship between two measurable quantities, $\delta\sbr{g}$ and
$\mathbf{v}(\mathbf{r})$, in terms of one unknown, $\beta
=\Omega\sbr{m}^{0.6}/b$.  The velocity predictions can then be compared to
measured velocities obtained through secondary distance indicators to
determine the quantity $\beta$.

In practice, one can use Equation (\ref{peculiar}) to determine $\beta
$ by several methods. A direct comparison of the well-known Local
Group (LG) velocity with the predicted LG velocity, $\mathbf{v}(0)$,
is one possibility.  However, in practice, for any redshift survey the
integral in Equation (\ref{peculiar}) is limited at some distance
$R\sbr{max}$. Thus determinations of $\beta$ by this method depend on
the assumptions about the convergence of the dipole at the survey
limit.

If we approximate the contributions from beyond $R\sbr{max}$ as a dipole
$\mathbf{U}$, then we may express Equation (\ref{peculiar}) as
\begin{equation}
{\mathbf{v}}\left( {\mathbf{r}}\right) =\frac{\beta}{4\pi }\int_0^{R\sbr{max}}
d^3{\bf{r}}^{\prime
}\delta\sbr{g}\left( \mathbf{r}^{\prime }\right) \frac{\left( \mathbf{r}^{\prime
}-\mathbf{r}\right) }{\left| \mathbf{r}^{\prime }-\mathbf{r}\right| ^3}+\mathbf{U}.
\label{peculiar2}
\end{equation}
Clearly a degeneracy exists between $\beta$ and $\mathbf{U}$ for the
peculiar velocity of the LG. By using peculiar velocity data for many
objects, one can fit for both $\beta$ and $\mathbf{U}$.
Alternatively, by making v-v comparisons in the LG frame,
$\mathbf{v}\sbr{LG}(\mathbf{r})=\mathbf{v}(\mathbf{r})-\mathbf{v}(0)$,
then the dipole contribution $\mathbf{U}$ cancels out.  Essentially
what is being measured is infall into overdensities with known
$\delta\sbr{g}$.

It should be noted however, measured velocities are radial which means
that our predictions, although three dimensional, will be converted to
radial velocities. We chose to describe radial velocities in the LG
frame, and calculate them according to
\begin{equation}
u\sbr{LG}\left( \mathbf{r}\right) =[\mathbf{v(r)}-\mathbf{v}(0)] \cdot
\frac{\mathbf{r}}{%
\left| \mathbf{r}\right| }.  
\label{radial}
\end{equation}

In this paper, we have reconstructed the galaxy density field using
the Two Micron All Sky Survey (2MASS) redshift data. The sample is
selected, weighted and smoothed in order to make predictions of the
local peculiar velocity field.  We employ the VELMOD technique
(Willick et al. 1997) to perform a maximum likelihood analysis which
compares the redshift and distance estimates to the derived gravity
field. In Section \ref{2MASS} we give the details of the selection,
correction and reconstruction procedures that have been applied to the
redshift data.  Section \ref{vv} describes v-v comparisons for the SN
Ia, SBF and SFI peculiar velocity datasets.  In Section \ref{MORPH} we
explore how splitting the density field by morphological type affects
our results. We repeat the tests of Section \ref{vv} using the
morphologically-segregated fields to make our velocity predictions. In
Sections 5 and 6 we present a discussion of the results and show
comparisons to values in the current literature.

\section{2MASS\ Gravity Field\label{2MASS}}

We now turn our attention to the construction of gravity field from
the 2MASS redshift survey.  In what follows, we discuss the selection
of the sample, general features and completeness corrections applied
to the data. We discuss the properties of the luminosity function and
how it relates to the weighting scheme used to define the redshift
space density field. Finally, we discuss the technique used to
transform the data from redshift to real space.

\subsection{2MASS Data \label{FEATURES}}

The 2MASS dataset provides an all-sky view of the nearby galaxy
population in the $J$, $H$ and $K_s$ bands.  Near-infrared light has
several advantages: first, it samples the old stellar population, and
hence the bulk of the stellar mass, and second it is minimally
affected by dust in the Galactic Plane.  Details of how extended
sources are identified are given by Jarrett et al. (2000).  We use
2MASS data to define a $K_s$ magnitude limited sample of galaxies.
Redshifts and morphological type designations (de Vaucouleurs et
al. 1991) are drawn from the HyperLeda\footnote{leda.univ-lyon1.fr}
database.

The selection criteria used to define the sample are as follows:
\begin{enumerate}
\item  $K_{20}$ apparent magnitude $\leq 10.5$
\item  Galactic latitudes $\left| b\right| \geq 11.5^{\circ }$
\item  Distance $r \leq 6500$ km s$^{-1}$ .
\end{enumerate}

The $K_{20}$ apparent magnitudes are those magnitudes defined to be
within the same circular aperture for a $K_s$-band isophote of 20 mag
arcsec$^{-2}$. The $K_{20}$ apparent magnitude limit of 10.5 is chosen
to yield a high redshift completeness ($\sim 90$\%) for the sample.

The cut in galactic latitude is chosen to reduce the incompleteness
associated with galactic extinction and confusion due to the dust and
stars, which increases closer to the plane of the Milky Way
$(b=0^{\circ})$.

The final selection criterion limits the volume to $6500$ km s$^{-1}$.
Given our $K_s$-band magnitude limit, beyond this distance only $L^*$
or brighter galaxies appear in the sample, hence shot noise becomes
large. Furthermore, previous work (Rowan-Robinson et al. 2000;
Kocevski et al. 2004) has shown that there are few significant
attractors beyond 6000 km s$^{-1}$, at least until the Shapley
Concentration is reached at 15000 km s$^{-1}$.

After making the distance and latitude cuts, the final magnitude
limited 2MASS sample comprises $5032$ galaxies. For each object in the
sample, all of the required information is known (except for 152 with
unknown morphological type). We have also cross-referenced these
galaxies in HyperLeda to obtain group identifications (Garcia et
al. 1993). This was successful for $25\%$ of our sample, leaving 1815
galaxies with group designations. Note that HyperLeda grouping
information only extends to a velocity distance of 5500 km s$^{-1}$
and has not been revisited even though additional galaxies have been
added to the HyperLeda database.

\subsubsection{2MASS Redshift Completeness\label{completeness}}

The parent 2MASS $K_{20} < 10.5, |b|>11.5^{\circ}$ catalog has a
redshift completeness of $90\%$. The completeness is not uniform
across the sky, however, and therefore small corrections are needed.
Since the redshift data are compiled from public sources, there are no
clear boundaries.  We note, however, that redshift completeness is
likely affected by two selection biases: location of observatories in
the Northern or Southern Hemispheres, and extinction in the Zone of
Avoidance (ZoA) at low Galactic latitudes.  Note that while the sample
here is K-band selected, and hence is much less affected by
extinction, redshifts are typically obtained in the optical and so the
redshift completeness will still be affected by extinction.

\begin{figure*}[tp]
\centering
\begin{tabular}{cc} 
\includegraphics[angle=-90,scale=0.3]{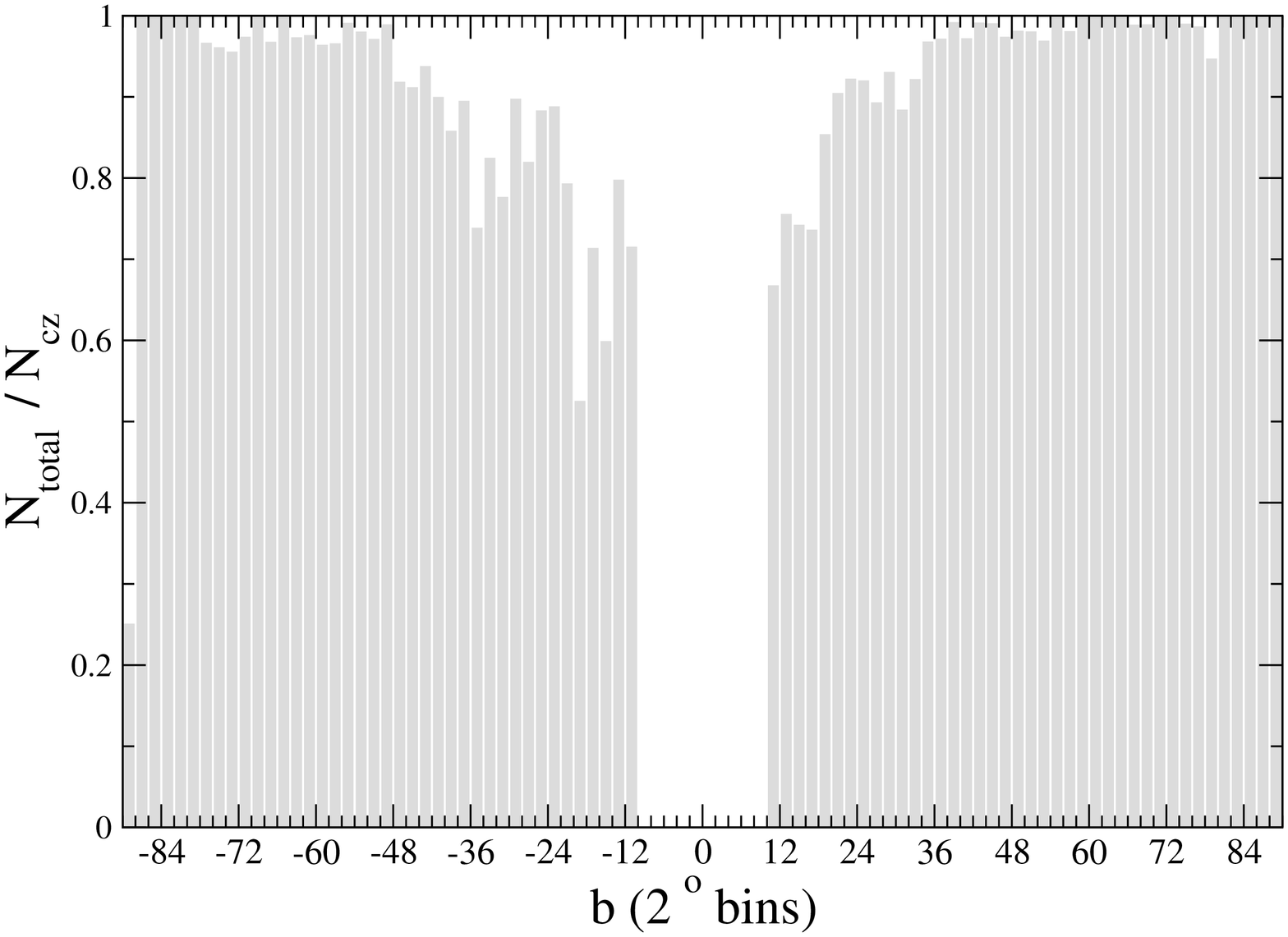}
\includegraphics[angle=-90,scale=0.3]{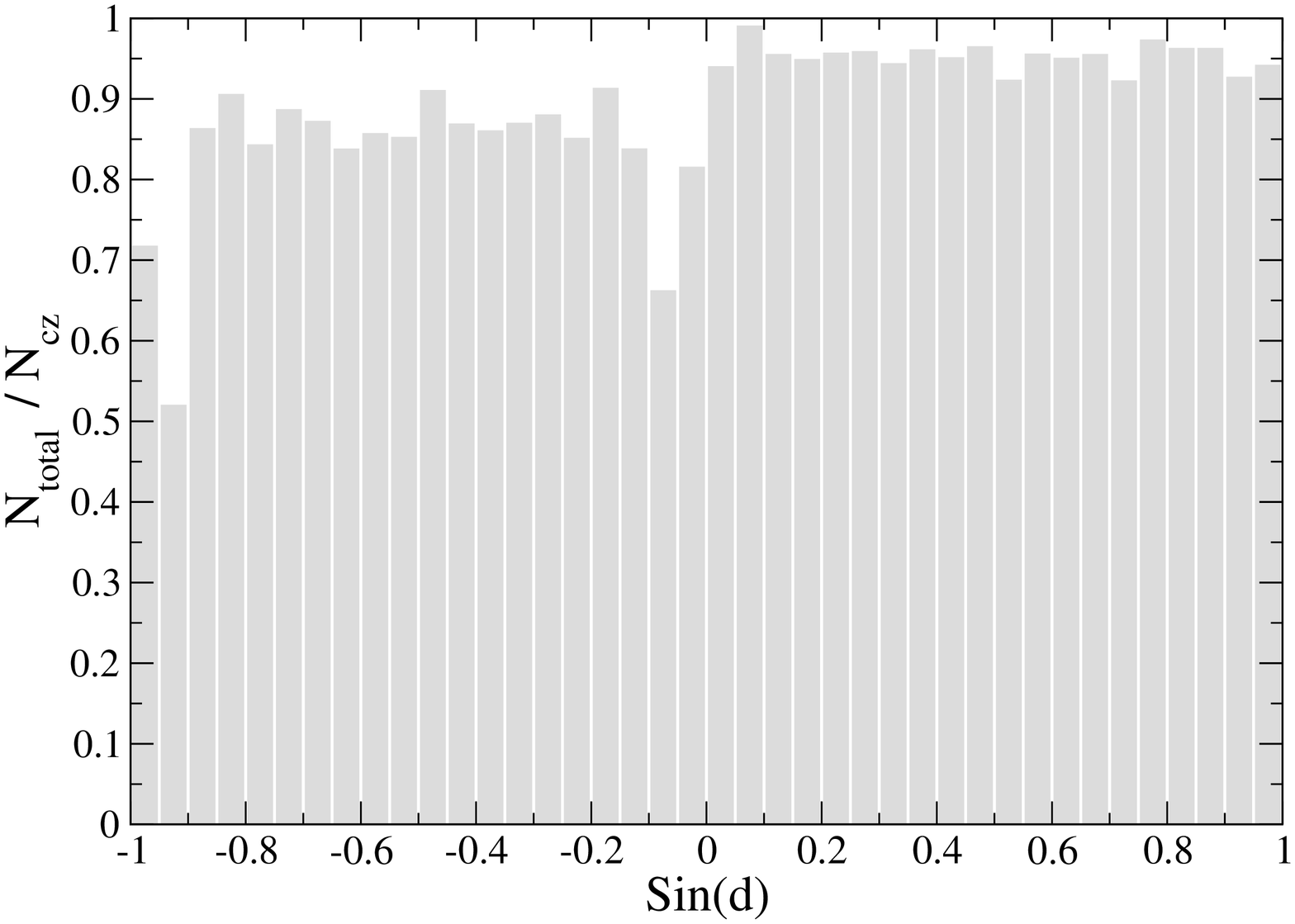}
\end{tabular}
\caption{{\it Left}: 2MASS redshift completeness is plotted as a
function of galactic latitude. The high latitude regions are very
complete, except for $b\backsim -90^{\circ }$ values where the
observed low value is due to a very small number of total galaxies. As
$b \rightarrow 0^{\circ }$ there is a decrease in the fraction of
measured redshifts due to the difficulty of obtaining redshifts close
to the galactic plane. The area $|b|<10^{\circ }$ is excluded. {\it
Right}: Redshift completeness in equal area bins of declination for
sin$|b|\geq 0.2$. Note the higher incompleteness in the southern
equatorial hemisphere.}
\label{fig1}
\end{figure*}

Fig. \ref{fig1} shows the 2MASS redshift completeness percentages in
terms of the galactic latitude and declination (D). While the
redshifts are highly complete at high Galactic latitudes, there is a
decline in the completeness at lower $|b|$. In the Northern celestial
hemisphere, the completeness is very high, but drops in the southern
hemisphere. We also find that the redshift incompleteness is not
correlated to the magnitude.

Since the overall completeness is high, and hence the corrections are
small, we model the incompleteness with a simple functional form. We
assume that the probability of observing a galaxy in our sample with a
redshift is a separable function $P_{cz}(b$,D)$ = P_1(b)P_2$(D).

Fig \ref{fig1} shows a completeness is $\sim 1$ at high Galactic
latitudes, with a decrease completeness towards the Galactic
plane. After trial and error, for $P_1(b)$ we adopt the functional
form
\begin{eqnarray}
P_1(b\geq 11.5^{\circ}) & = & 1-\exp \left[-\frac{\sin b}{\sin b\sbr{N}}\right] 
\nonumber \\
P_1(b\leq -11.5^{\circ}) & = & 1-\exp \left[-\frac{\sin b}{\sin b\sbr{S}}\right]
\end{eqnarray}

For declination, the northern hemisphere is quite complete whereas the
southern hemisphere is less so. Fig, \ref{fig1} shows no strong
dependence on declination within a given hemisphere, so we have chosen
to simply fit with a step function
\begin{eqnarray}
P_2(\mathrm{D} \geq 0) & = &  \mathrm{D}\sbr{N}\equiv{ 1}  \nonumber \\
P_2(\mathrm{D} \leq 0) & = & \mathrm{D}\sbr{S}
\end{eqnarray}

The likelihood of the given set of galaxies with and without redshift
is
\begin{equation}
{\cal{L}}_{cz} =\prod_{i=1}^{N_{cz}}P_1(b_i)P_2(\mathrm{D}_i) \times
\prod_{i=1}^{N\sbr{miss}}1-[P_1(b_i)P_2(\mathrm{D}_i)]
\end{equation}
where the first product sum is over galaxies with measured redshift
and the second is over those having been detected with no associated
redshift. 

The maximum likelihood solution yields best fit parameters $b\sbr{N}=
8.33^{\circ}$ and $b\sbr{S}=-11.9^{\circ}$, and D$\sbr{N}\equiv1$ and
D$\sbr{S}=0.933$.  We correct for incompleteness in $b$ and
declination by weighting the galaxies with redshifts by
$1/[P_1(b)\,P_2$(D)].

\subsubsection{Cloning the Zone of Avoidance\label{clone}}

Due to increasing redshift incompleteness at low galactic latitudes,
only galaxies satisfying $|b| \geq 11.5^{\circ }$ were
selected. Leaving the unsurveyed regions empty would cause systematic
errors in the dynamical model, because this region would behave as a
void.  This would have the effect of creating a false outflow, as can
be understood through the $\delta $ term in Equation (\ref{peculiar}).

There are several ways in which this effect can be corrected.  For
instance, the volume can be filled with a uniform distribution of
galaxies that exactly match the surveys' average density. This method
has the benefit of alleviating the systematic errors associated with
the empty space, and since $\delta =0$ in this region, there is no
gravitational effect.

We chose instead to fill the plane by interpolating or {\it cloning}
adjacent, equal area regions above and below the galactic equator
(Hudson 1993). Due to the infrared nature of the survey, a galactic
latitude cut as low as $|b|=11.5^{\circ }$ could be made, which meant
only a relatively small strip had to be cloned, as opposed to optical
surveys which typically have higher limiting $|b|$ (Santiago et
al. 1995; Giuricin et al. 2000). Hudson (1994) found that if uniform
density was assumed then the values of $\beta$ are only $\sim 10 \%$
higher.  However, surveys in the ZoA (Staveley-Smith et al. 2000)
suggest that cloning is a better approximation.

In our cloning procedure, the $K_{20}$ apparent magnitude and
redshift, along with all of the other associated properties of a
cloned galaxy remain the same as the parent, with the exception of a
new galactic latitude which depends on the parent and cutoff galactic
latitudes through, $
\sin{b_{\mbox{\begin{footnotesize}clone\end{footnotesize}}}}=
\sin{b_{\mbox{\begin{footnotesize}parent\end{footnotesize}}}}-\sin{b\sbr{min}}
$.

\subsection{Luminosity Function: $\Phi \left( L\right) $ \label{LUMINOSITY}}

In order to correct for galaxy incompleteness due to the imposed flux
limit, it is necessary to know the luminosity function (hereafter LF).
The empirical LF is often fitted by an analytical expression first
described by Schechter (1976),
\begin{equation}
\Phi \left( L\right) dL=\phi ^{*} \left(\frac{L}{L^{*}}\right)^\alpha\exp\left[-\frac{L}{L^*}\right] \frac{dL}{L^{*}}  
\label{LF}
\end{equation}
in which $L^{*}$ (or equivalently $M^{*}$ if we describe it in terms
of absolute magnitudes) is the fiducial magnitude and represents the
point in which the LF changes from a power law (with slope $\alpha $)
to an exponential.  

To estimate the LF parameters, we adopt the density-independent method
of Sandage et al.  (1979).  In Section \ref{procedure} below, we will
outline the redshift transformation method which uses the smoothed,
weighted density field to make iterative corrections to the distance
of each galaxy based on an assumed $\beta$.  After each iteration,
when the distances to every galaxy have been updated, we recalculate
the LF parameters.  Thus in contrast to, for example, deep samples in
which redshift is used as a proxy for distance, in our case a
self-consistent set of LF parameters for each value of $\beta$ is
derived.  Nevertheless, at low redshifts the uncertainty in a galaxy's
distance due to its peculiar velocity is large, so we have used only
galaxies in the distance range $1500$ km s$^{-1}$ to $6500$ km
s$^{-1}$ to estimate the LF parameters, and excluded faint galaxies
with $M_{K_{20}} > -18$.

We obtain best fitting values $M^{*}_{K_{20}}=-23.42$, $\alpha=-0.982$
and $\phi^{*}_{K_{20}}=0.011$ (in $h^3$Mpc$^{-3}$) from the analysis
of the real space galaxy distribution (at the best fitting $\beta$,
see Section \ref{BETARESULTS}).  This is in agreement with the results
of Cole et al. (2001) who used the 2MASS and 2dF galaxy redshift
surveys to derive values for the parameters $M^{*}_{K_{20}}-5$log${h}$
, $\alpha $ and $\phi^{*}_{K_{20}}$ in the $K_s$ band yielding,
$-23.44 \pm 0.03$, $-0.96 \pm 0.05$ and $0.0108 \pm 0.0016 h^3$
Mpc$^{-1}$ respectively.  The likelihood analysis of the LF was
performed with and without the inclusion of cloned galaxies.  We find
that there is no appreciable difference in either case, as expected
since the Sandage et al. method is density-independent.  In Section
\ref{MORPH} we explore how morphology affects the LF parameters by
segregating the density field into early and late-types.

\subsection{Transforming Redshifts to Real Space Distances \label{R2R}}

\subsubsection{Weighting the Sample\label{weight}}

Knowing the LF and completeness of the sample, we can calculate the
selection function (the probability that a galaxy is included in the
redshift sample) at any distance $r$ as,
\begin{eqnarray}
\phi \left( r\right) & 
\equiv &
\frac{\int_{4\pi r^2f_{\min }}^\infty \Phi
\left( L\right) dL}{\int_{L\sbr{min}}^\infty \Phi \left( L\right) dL%
}\qquad r\geq r\sbr{min}  \label{selection} \\
&\equiv &1\qquad \qquad \qquad \qquad \;\;r<r\sbr{min}  \nonumber
\end{eqnarray}
where $f\sbr{min}$ is the flux corresponding to $K\sbr{20,lim}=10.5$.
In practice the luminosity function is poorly constrained at the faint
end, and hence we place a lower limit on the integral at $L\sbr{min} =
4\pi r\sbr{min}^2f\sbr{min}$. $L\sbr{min}$ corresponds to $M_K =
K\sbr{20,lim}-5 \log(r\sbr{min}/10 \mathrm{pc})$, which is $-18$ for
$r\sbr{min} = 5 h^{-1}$ Mpc.

We assign extra weight, $1/\phi(r)$, to galaxies with measured
redshifts to account for the galaxies in the same volume that were not
detected or failed to meet the selection criteria. This is equivalent
to adding galaxies at every position where we have a galaxy with a
measured redshift.

Weights are assigned to all galaxies in our sample limit, $r \leq
R\sbr{max}$.  Galaxies outside of $R\sbr{max}$ are set to a weight of
zero. The sum of the weights divided by the volume within $R\sbr{max}$
yields the average number density.

\subsubsection{Smoothing the Sample \label{smoothing}}

The galaxy distribution obtained from redshift surveys is a point
process.  If we were to apply linear theory directly, the predictions
would diverge near the weighted points.  We smooth the density field
so that the velocity field is continuous and so that linear theory
applies. This is accomplished by applying the following equation to
the density field,
\begin{equation}
\delta \left( \mathbf{r}\right) =%
\frac{\rho \left( \mathbf{r}\right)-%
\overline{\rho }}{\overline{\rho}} =%
\frac {1}{\overline{n}}\sum_i^{N\sbr{gal}}\frac{1}{\phi(r_i)}
\left[W\left( 
\frac{\left| \mathbf{r-r}_i\right| }{r\sbr{sm}}\right) -1 \right]
\end{equation}
in which $W$ is a smoothing kernel and $\overline{n}$ is the number
density of the $N\sbr{gal}$ weighted objects.  We have chosen to use a
top-hat kernel in which $W=|\mathbf{r}-\mathbf{r}_i|^3/r\sbr{sm}^3$ for
$\left|\mathbf{r}-\mathbf{r}_i\right| <r\sbr{sm}$ and set to unity
otherwise.  We use a fixed smoothing length $r\sbr{sm}$ of $500$ km
s$^{-1}$ independent of distance.

\subsubsection{Reconstruction Procedure\label{procedure}}

We now discuss the reconstruction method used to transform the
measured redshifts into real space distances, $r$, for the
model. Given a smoothed all sky density field, the peculiar velocity
of any galaxy can be estimated via Equation (\ref{peculiar2}), and
used to make distance corrections to the sample. We follow a similar
method to that of Yahil et al. (1991) who use an iterative technique
in which gravity is adiabatically ``turned on'' by increasing $\beta$.
The following outlines a recipe for a self-consistent solution to real
space density and velocity maps given only positions and redshifts.

Prior to the start of the iterative procedure, we collapse the
so-called ``Fingers of God'' (nonlinear redshifts distortions
resulting in an apparent radial stretching along the line of sight) by
placing all grouped galaxies at a common median redshift,
$cz\sbr{LG}$, and angular position.

Initially, galaxies are assigned distances that are equal to their
Local-Group-frame redshifts.  A value of $\beta$ is defined at each
step of the iteration and the resulting peculiar velocity field is
calculated.  We perform 100 steps, defining a density and velocity
field at each iteration from $\beta =0.01$ to $\beta =1.00$ in steps
of $0.01$.  At each iteration the following steps are performed.

\begin{enumerate}

\item The likelihood function for the LF is minimized to determine the
best fit parameters and a number weight is assigned to each galaxy
according to its distance as described in Section \ref{weight}.

\item Galaxies are cloned to fill the void as described in Section
\ref{clone}, assigning each cloned galaxy the properties of its
parent, including its weight.

\item The average density $\overline{n}$ of all galaxies within
$R\sbr{max}$ is calculated.  This then defines the density contrasts
$\delta\sbr{g}$.

\item A peculiar velocity is calculated for each galaxy using Equation
(\ref{peculiar2}). We average the current peculiar velocity with that
of the past 5 iterations. As we step through $\beta$, we are
essentially turning on the gravity, and since increases in $\beta $
are very small at each step, changes in weight and positions are also
small over this range. The averaging has the benefit of damping out
any unphysical oscillatory behavior

With this information we update the distance of each galaxy according
to the equation for total recession velocity,
$r_i=cz\sbr{LG,i}-u(r_i)$ (where $i$ denotes the $i^{th}$ of
$N\sbr{gal}$ galaxies). Note that the distance to each galaxy in the
sample changes at each iteration. In particular, after an iteration, a
galaxy's distance may ``move'' it across $R\sbr{max}$. Therefore, at
each iteration, we update the positions of all galaxies with
$cz\sbr{\mathrm LG} < 9000$ km s$^{-1}$, which is well beyond
$R\sbr{max}$. A galaxy residing in the region $r > R\sbr{max}$ and
$cz_{\mathrm LG} < 9000$ km s$^{-1}$ is assigned a weight of zero, and
therefore contributes nothing to the density and predicted peculiar
velocity fields. If a subsequent iteration brings it within
$R\sbr{max}$, it is assigned a weight. Similarly, if an iteration
brings a galaxy from $r < R\sbr{max}$ to $r > R\sbr{max}$ it is
assigned a weight of zero. In this way, we maintain a density field
based on a self-consistent distance limit $R\sbr{max}$.

\end{enumerate}

We repeat the above steps at each iteration.  Thus after the iterative
procedure is complete, we have 100 density and velocity fields
(specified by $\beta$), each of which can be compared to measured
peculiar velocity data.

\subsubsection{Cosmography \label{Cosmography}}

\begin{figure}[tp]
\centering
\plotone{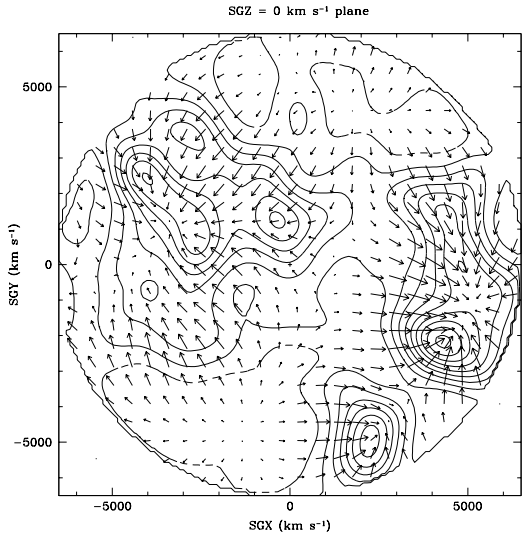}
\caption{The isodensity contours for a slice through the Supergalactic
Plane of the reconstructed 2MASS density field.  The density field has
been smoothed with a Gaussian kernel of $500$ km s$^{-1}$. The heavy
contour indicates mean density, $\delta \sbr{g}=0$, and increases in steps
of $\Delta \delta\sbr{g} = 0.5$. It can be seen that the zone of avoidance
has been filled with cloned galaxies. Superimposed is the predicted
peculiar velocity field for $\beta_{K} = 0.49$. }
\label{fig2}
\end{figure}

Fig. \ref{fig2} shows the Supergalactic Plane maps of the
reconstructed density field and the resulting velocity field for our
best fit $\beta_{K} = 0.49$ as derived using the 2MASS redshift
data. The galaxy density field is smoothed using a Gaussian kernel
with a smoothing radius of $500$ km s$^{-1}$.

The plot shows the major structures that are located in the
Supergalactic SGX-SGY Plane as well as the resulting flow
fields. Directly above the center (LG) in the positive SGY direction
is the Virgo Supercluster, with an associated peak overdensity of
$\delta\sbr{g}=2.7$. At $SGX = -4000$ to $-3000$ km s$^{-1}$ and $SGY
= 1000$ to $2500$ km s$^{-1}$ is the Great Attractor (GA). The peak in
seen in Fig. \ref{fig2} is coincident with the cluster Abell 3574
($r\sim 5300; \delta\sbr{g} = 3.4$), but the highest peak in this
region lies slightly off the Supergalactic Plane, and is coincident
with the Centaurus cluster at ($r \sim 3600$ km s$^{-1}$;
$\delta\sbr{g} = 4.7$) with a secondary peak at Pavo II (Abell S0805;
$r \sim 4500$ km s$^{-1}$; $\delta\sbr{g} = 4.6$).  Note that what may
be the most massive cluster in the GA region, Abell 3627
(Kraan-Korteweg et al. 1996) is not in our sample because of its low
Galactic latitude ($b = -7^{\circ}$).  However, the dynamical
influence of a single cluster, even a massive one such as Abell 3627,
is small.  Instead the peculiar velocity field is more strongly
affected by the large-scale overdensity of the supercluster as a
whole. It is these large-scale overdensities that are crudely
reproduced by our cloning procedure.

The Perseus-Pisces (PP) supercluster shows the largest overdensity of
$\delta\sbr{g}=5.5$, coincident with the Perseus cluster, located near
the sample limit and at the edge of the ZoA, in the SGX, -SGY
direction. Note however that the overdensity may be enhanced due to
the cloning of Perseus. The region at $SGX = 2200$ km s$^{-1}$ and
$SGY = -5000$ km s$^{-1}$ is harder to classify. It has an overdensity
of $\delta\sbr{g}=2.6$, and appears to coincide with Abell 168A/Abell
1194 overdensities. The Coma cluster lies outside of our sampled
distance limit at around a redshift of $cz=7000$ km s$^{-1}$. In
contrast, the -SGX, -SGY quadrant is almost completely devoid of
overdensities, and is dominated by the Sculptor void which exhibits a
weak velocity outflow.

\section{Velocity-Velocity Comparisons\label{vv}}

With the full velocity field modelled, we now make comparisons to
peculiar velocity data sets. Below is a discussion of the properties
of the peculiar velocity sets used, as well as the methods used in
their comparison to the model.

\subsection{Peculiar Velocity Data \label{PECSETS}}

We compare our predictions to three published peculiar velocity data
sets.  Each of which vary in sample size and typical distance errors
and method of obtaining the secondary distance information.

Our subsample of the Spiral Field I-Band (hereafter SFI; Haynes et al
1999a,b) survey consists of 836 galaxies of morphological type Sbc-Sc,
in which distances have been derived from the I-Band Tully-Fisher
relation over the full sky. Typical distance errors are on the order
of 20\% for our subsample, which extends to $5000$ km s$^{-1}$. The
characteristic depth of the sample, which is the weighted distance
from which which the majority of the signal arises is
\begin{equation}
d_c = \frac{\sum d_i w_i}{\sum w_i}
\end{equation}
where the weights $w_i = 1/\sigma_{d,i}$ and $\sigma_{d,i}$ is the distance
error. For the SFI set, $d_c= 2700$ km s$^{-1}$, so the characteristic distance
error is $\sigma_{d_c}= 540$ km s$^{-1}$.

The I-band Surface Brightness Fluctuation (SBF, Tonry et al. 2001)
survey consists of 266 galaxies extending to a distance of ~$4000$ km
s$^{-1}$. The uncertainties in the distance estimates varies inversely
with the resolution of the images. Typical distance errors for our
sample are of the order of 8\%, approximately half that of SFI. The
characteristic depth is 1200 km s$^{-1}$ having $\sigma_{r_c}= 96$ km
s$^{-1}$.

The final data set is a compilation of supernovae of type Ia (SN Ia,
Tonry et al. 2003) and consists of 59 SNe extending to a distance of
$6000$ km s$^{-1}$, with typical errors of 8\%. The characteristic
depth of this sample is 2200 km s$^{-1}$ with $\sigma_{r_c}= 176$ km
s$^{-1}$.

\subsection{Method \label{VELMOD}}

We apply two v-v comparison methods. Our primary method is a slight
variant of the VELMOD method. For comparison, we also apply a simple
and complementary $\chi^2$ fitting method.

\subsubsection{VELMOD}

The procedure outlined here differs slightly from the VELMOD method by
Willick et al. (98).  Our procedure uses two free parameters: $\beta$,
which scales the predicted peculiar velocities, and $H_0$, which
allows for a rescaling of the published distances. In contrast,
Willick et al. fit $\beta$ as well as the parameters of the TF
relation.

Specifically, we maximize the probability of each galaxy having its
observed velocity, $P\left( cz\sbr{obs}\right)$. We construct the joint
probability distribution of redshift and the (unobservable) true
distance,
\begin{equation}
P\left( cz\sbr{obs},r\right)= P\left(cz\sbr{obs}|r\right)P\left( r\right)
\label{P(cz,r)}
\end{equation}
where the first term 
\begin{equation}
P\left( cz\sbr{obs}|r\right) =\frac 1{\sqrt{2\pi }\sigma _v}\exp 
\left[ - \frac{\left[
cz\sbr{obs}-\left( H_0r+\beta u\left( r\right) \right) \right] ^2}{2\sigma _v^2}%
\right]
\label{P(cz|r)}
\end{equation}
is a description of the redshift-distance relation in terms of a
Gaussian probability distribution. The predicted $cz\sbr{pred}$ has three
parts; an expansion velocity; a part due to the linear perturbations
of the surrounding overdensities, $u(r)$, and a part which is often
referred to as {\it velocity noise}, $\sigma_v$, assumed to be
Gaussian, arising from strongly non-linear processes.  For this
calculation, $\sigma_v$ is fixed at $200$ km s$^{-1}$ for field
galaxies. This allows for uncertainties in the linear predictions,
nonlinear motions and observational errors in $cz\sbr{obs}$.  This value
of $\sigma_v$ is chosen because it gives reasonable reduced $\chi^2$
fits (see Section \ref{Chi}), although our recovered values of $\beta$
are not sensitive to precise value chosen for $\sigma_v$.  An
additional velocity error was added in quadrature to $\sigma_v$ to
account for extra velocity dispersion for galaxies associated with
clusters: $\sigma\sbr{cl} = \sigma_0/\sqrt{1+(r/r_0)^2}$, where
$\sigma_0=700(400)$ km s$^{-1}$ and $r_0= 2(1)$ Mpc are adopted for
Virgo (Fornax) respectively (Blakeslee et al. 1999).

The second term in eq. \ref{P(cz,r)} is the a priori probability of
observing an object at distance $r$
\begin{equation}
P\left(r\right) \propto \exp \left[-\frac{\left(r-d \right)
^2}{2\sigma_d^2}\right]\left(1+\delta\sbr{g} \right)
\label{P(r)}
\end{equation}
and is a Gaussian with mean equal to the estimated secondary distance
$d$ with an additional term $(1+\delta \sbr{g})$, proportional to the
number density of galaxies, to account for inhomogeneous Malmquist
bias. The $(1+\delta \sbr{g})$ term is evaluated using the 2MASS density
field, smoothed with a Gaussian filter of $500$ km s$^{-1}$. In
principle, for each peculiar velocity dataset, one should use
$\delta\sbr{g}$ for the appropriate morphological type, for example,
$\delta\sbr{E}$ for the SBF dataset. In practice, as we will discuss in
Section \ref{Discussion}, the correction arising from the
$(1+\delta\sbr{g})$ term is small except for the SFI case. In Section
\ref{MORPH} we show that the bias $b$ for Spiral and later types
appears to be very similar to that of the sample as a whole, thus
justifying the simpler approach adopted here.

The expression $P(cz\sbr{obs})$ can be obtained by integrating over all
possible distances,
\begin{equation}
P\left( cz\sbr{obs}\right) =\int_0^\infty P\left( cz\sbr{obs}|r\right) P\left(
r\right)
dr \,.
\label{probcz}
\end{equation}
Equation (\ref{probcz}) is then normalized over all possible
velocities.

The VELMOD method then maximizes the log likelihood, $\sum_i
\ln[P(cz_i)]$, over all galaxies in the peculiar velocity data set as
a function of the free parameters $\beta$ and $H_0$.

\subsubsection{$\chi^2$ method \label{Chi}}

In addition to VELMOD, we perform a simple $\chi^2$ minimization.  The
$\chi^2 $ function is constructed from the differences between the
observed $cz\sbr{obs}$ and that which is predicted from a combination of
our radial velocity and the galaxies measured distance. We minimize
the function,

\begin{equation}
\chi^2 =\sum \frac{[cz\sbr{obs}-(H_0 d+\beta
u(d))]^2}{\sigma_{cz}^2}
\label{chi}
\end{equation}
which is similar to that contained inside the parenthesis of Equation
(\ref{P(cz|r)}).  In this case however we do not evaluate $u(r)$ at
all possible values of $r$ and marginalize, but instead use the
peculiar velocity at the estimated distance $d$.  To the extent that
the peculiar velocity predictions do \emph{not} change rapidly on the
scale of the distance error, $\sigma_r$, this approximation will be
accurate.  The error $\sigma_{cz} = \sqrt{{\sigma_v}^2 +
{\sigma_d}^2}$, where $\sigma_v$ is the same as in Equation
(\ref{P(cz|r)}). Note that this application of the $\chi^2$ method
neglects inhomogeneous Malmquist bias. Although it is possible to
correct for the estimated distances for this bias (Hudson 1994a), we
have not done so in this paper.

\subsection{Results\label{BETARESULTS}}

\begin{figure}[tp]
\centering
\includegraphics[width=\columnwidth]{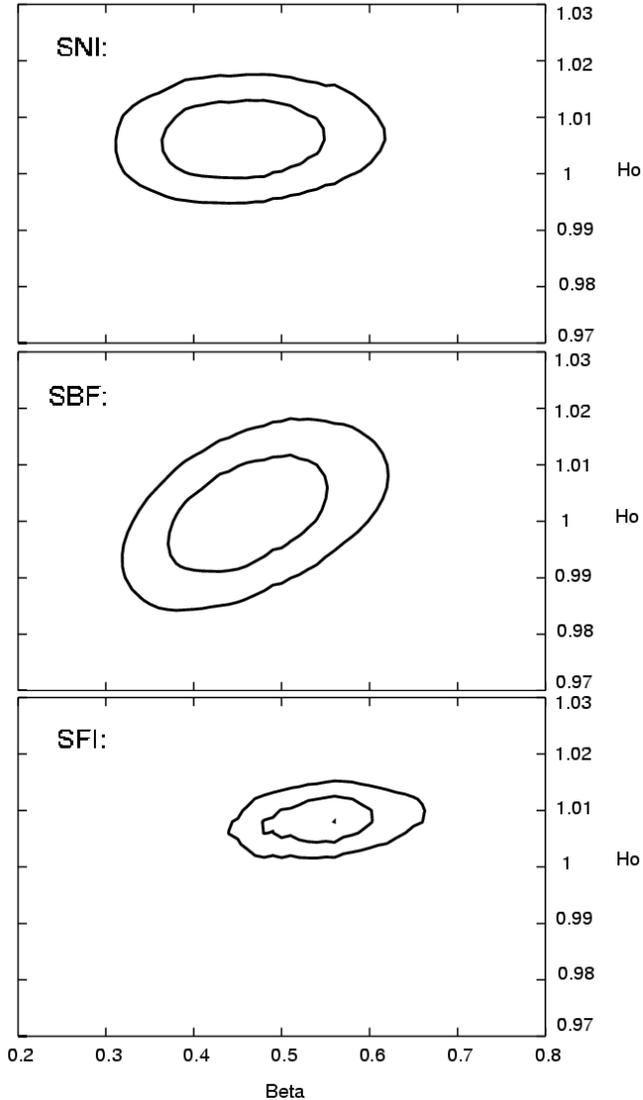}
\caption{The $68\%$ and $90\%$ joint probability contours on $\beta$
and $H_0$ are plotted for the VELMOD likelihood analysis for each
peculiar velocity data set (SN Ia, SBF, SFI) and the derived 2MASS
velocity field. Note that $H_0$ is very close to unity in all
cases. This joint distribution is then marginalized (over $H_0$) to
obtain a probability distribution function for $\beta$.}
\label{fig3}
\end{figure}

The VELMOD analysis applied to the SN Ia, SBF and SFI peculiar
velocity sets yields the likelihood confidence ellipses shown in
Fig. \ref{fig3}.  Table \ref{tbl-1} summarizes the results for the
peculiar velocity comparisons with those predicted by the
reconstructed 2MASS density field. $\beta\sbr{VELMOD}$ indicates the
value obtained from the first moment of the marginalized $\beta$
probability distribution function (PDF) obtained from the VELMOD
analysis. The $\beta\sbr{VELMOD}$ values are then combined to determine
the best fitting $\beta=0.49 \pm 0.04$, which is indicated as {\it
All} in the last row of Table \ref{tbl-1}.  The $\beta_{\chi^2}$
values are quoted for the $\chi^2$ minimization. There was only one
free parameter in the $\chi^2$ fits corresponding to $\beta$, and
$H_0$ was held constant at the best fit VELMOD value. Since there is
only one free parameter in the fits, the number of degrees of freedom
(DOF) is $N\sbr{gal}-1$.

\begin{table}[tp]
\begin{center}

\caption{Summary of $\beta$ determinations: VELMOD and $\chi^2$
minimization results.}

\begin{tabular}{lcccc} 
\hline
\hline
Set:  & $\beta\sbr{VELMOD}$  & $\beta_{\chi^2}$ & D.O.F &  $\chi^2$\\ 
\hline
SN Ia & $0.47 \pm 0.06$ & $0.45 \pm ^{+0.09}_{-0.05}$ &$59$   & $42$ \\ 
SBF   & $0.46 \pm 0.06$ & $0.44 \pm ^{+0.05}_{-0.07}$ &$266$  & $238$ \\
SFI   & $0.55 \pm 0.05$ & $0.53 \pm ^{+0.04}_{-0.03}$ &$831$  & $756$ \\
All  & $0.49 \pm 0.04$ & $-$  & $-$  & $-$ \\ 
\hline
\end{tabular}
\label{tbl-1}
\end{center}
\end{table}

\begin{figure}[tp]
\centering
\includegraphics[width=0.9\columnwidth]{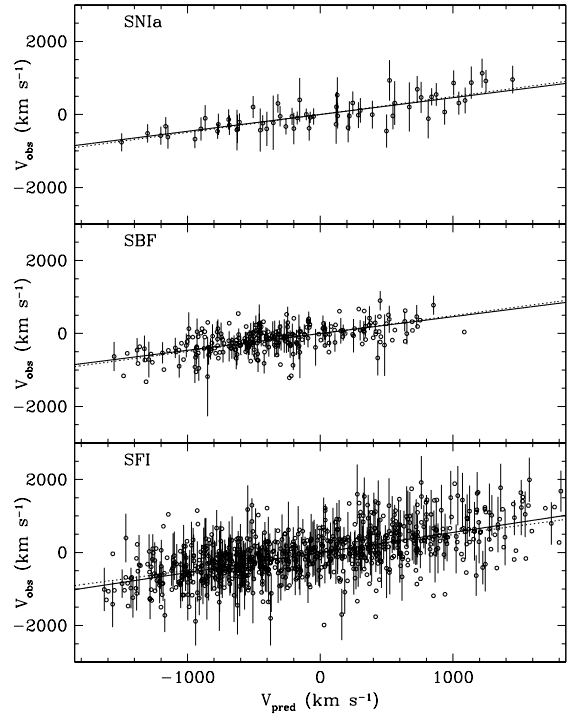}
\caption{Comparisons of SN Ia, SBF and SFI peculiar velocities to
those which have been derived from the 2MASS modelled velocity field
out to a distance of $6500$ km $s^{-1}$.  Errors are defined by
$\sigma_{cz}$, but for SFI and SBF data only one in three error bars
are plotted.  Predictions are scaled to $\beta = 1$, so the slope of
best fit corresponds to the value of $\beta$. The solid line indicates
the best fit to the individual data set shown, and the dotted line is
the global fit ($\beta=0.49$).  }
\label{fig4}
\end{figure}

With the adopted value of $\sigma_v$, the $\chi^2$ values are
reasonable.  The agreement between the SN Ia and the SBF samples is
evident, and these two samples agree (within errors) with the slightly
higher $\beta$ determination for the SFI sample.  All peculiar
velocity datasets agree within the determined errors. The agreement is
reinforced in Fig. \ref{fig4}, where we plot the measured SN Ia, SBF
and SFI peculiar velocities against those predicted by 2MASS for each
respective $\beta\sbr{VELMOD}$.

\section{Effect of Morphology
\label{MORPH}}

It is likely that mass is related to light in a way more complicated
than suggested by linear biasing.  While such deviations may be small
on linear scales probed by peculiar velocities, there remains the
possibility of measuring such deviations. In particular, it is
well-known that early-type galaxies are preferentially found in dense
environments.  In the case of a toy model in which, for example, all
mass of the Universe was in clusters, with no mass in the field, then
one would expect the elliptical density field to be a better predictor
of peculiar velocities than that of spirals. So it is of interest to
explore the predictions of density fields pre-selected by
morphological type.

We have morphological types for nearly $97\%$ of our sample.  We
consider two subsamples, an early-type (E+S0) sample having $T\leq0$,
and a late-type (S+Irr) sample having $T>0$. Here T represents a
numerical code for the revised (de Vaucouleurs) morphological
type. Our cloned sample contains $N\sbr{E+S0}=2486$ galaxies. Despite the
$K_s$-band selection, the 2MASS redshift survey contains more
late-type galaxies than early-types, such that $N\sbr{S}=3795$.

We analyzed both in the same manner as outlined in Section
\ref{procedure}.  The LF parameters for the early-type galaxies are
$M^{*}_{K_{20}}=-23.57$ and $\alpha=-0.72$ and the corresponding
values for the late-types are $M^{*}_{K_{20}}=-23.08$ and
$\alpha=-0.79$. Note that both of these are flatter than the full
density field, which may be a result of incompleteness in
morphological types at the faint end. This trend is also observed in
the recent determination of early and late-type LF's by Kochanek et
al. (2001). They find $M^{*}_{K_{20}}=-23.53\pm0.06$ and
$\alpha=-0.92\pm0.10$ and $M^{*}_{K_{20}}=-22.98\pm0.06$ and
$\alpha=-0.87\pm0.09$ for the early and late-type LF parameters
respectively.  They used $7\leq K_{20}\leq11.25$ mag, with $cz>2000$km
s$^{-1}$ and a limiting latitude of $|b|\geq 20^{\circ }$ for an
approximately equal sample size.

\begin{figure*}[tp]
\centering
\plottwo{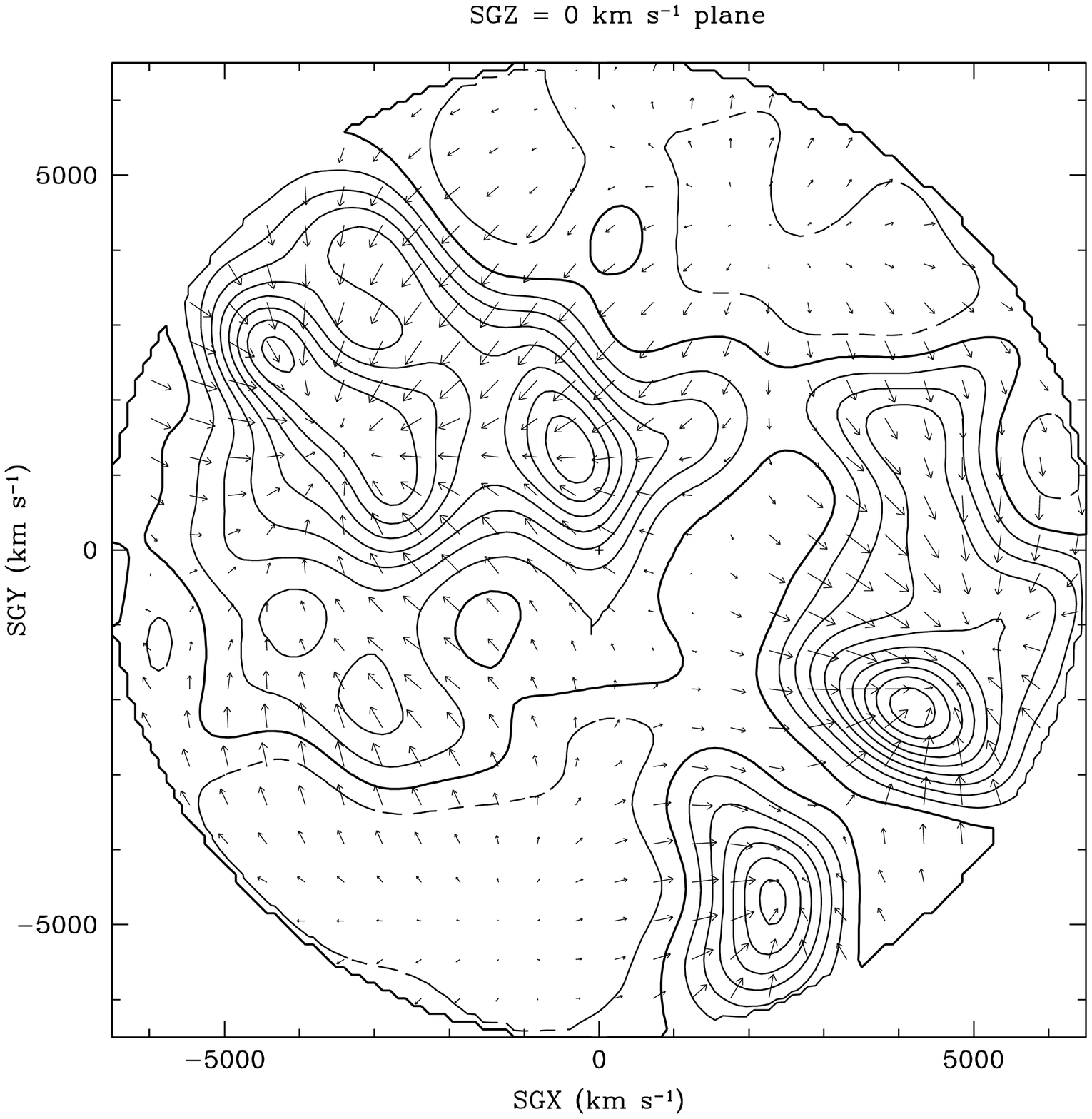}{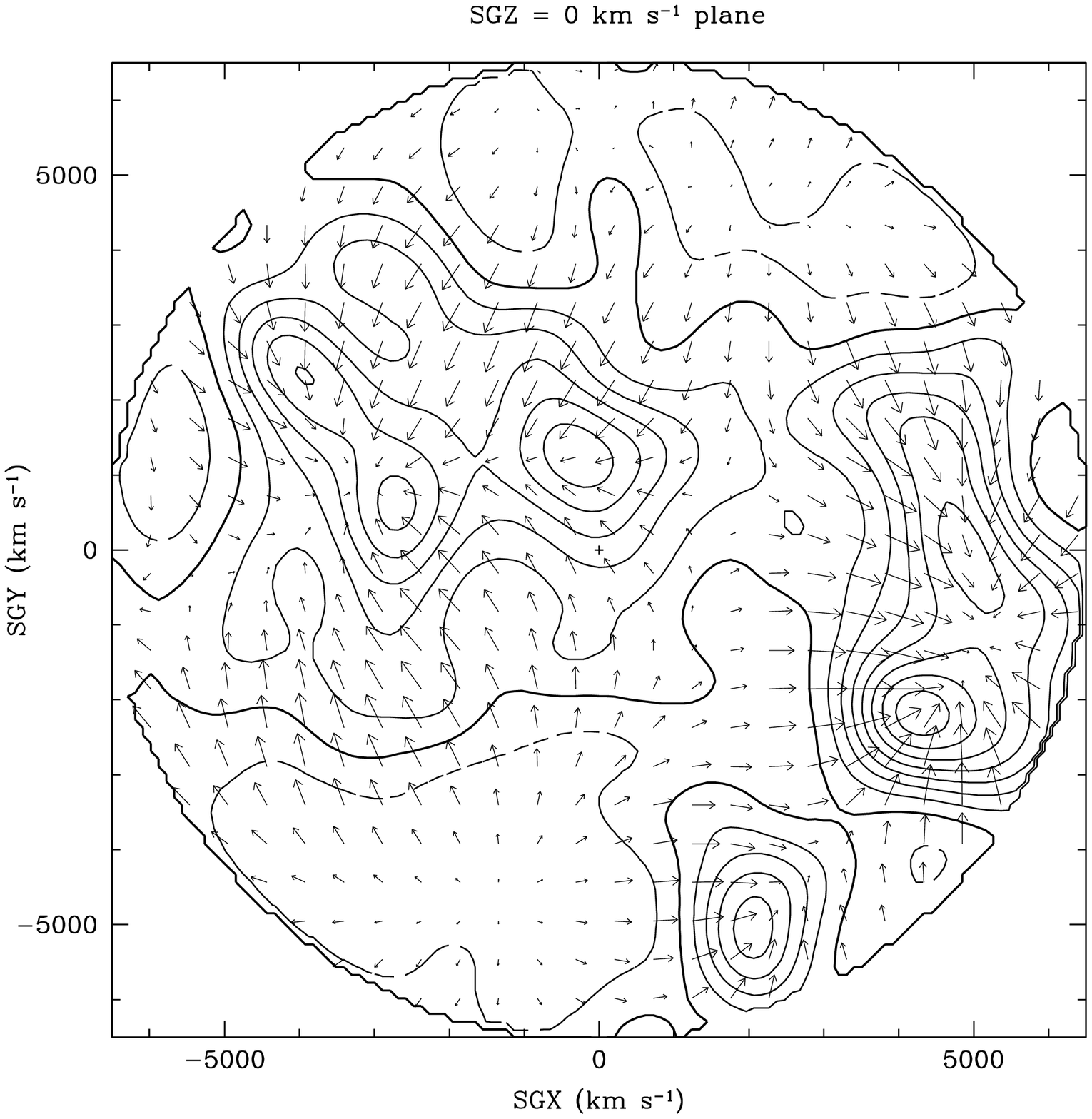}
\caption{ The reconstructed E+S0 ({\it left}) and S+Irr ({\it right})
density and velocity fields. Contours are as in
Fig.\ref{fig2}. Predicted peculiar velocities are for the best fit
values from Table 2. While general features are consistent in both
fields, the relative contrast indicates that the early-type galaxies
are more clustered.  }
\label{fig5}
\end{figure*}

The derived density and velocity fields from the two morphological
subsamples are shown in Fig. \ref{fig5}.  The most distinct difference
between the fields is that the early-types clearly reside in higher
density regions. The density contrast is more significant in every
clustered region, as is indicated by the increase in contour line
density.  Since the amount of clustering is quantified through the
biasing parameter $b$, we can also measure $b$ via the $\beta$ of best
fit to the peculiar velocities.

\begin{table}[tp]
\begin{center}
\caption{ $\beta$ dependency on morphology.}   

\begin{tabular}{lcccc} 
\hline
\hline
Set:  & $\beta\sbr{VELMOD}$  & $\beta_{\chi^2}$ & D.O.F &  $\chi^2$\\ 
\hline 
&& Late-Type (Spirals)& &\\
\hline
SN Ia & $0.46 \pm ^{+0.06}_{-0.07}$ & $0.48 \pm ^{+0.05}_{-0.07}$ &$58$   & $46$ \\ 
SBF   & $0.45 \pm 0.06$ & $0.38 \pm ^{+0.06}_{-0.04}$ &$265$  & $244$ \\ 
SFI   & $0.50 \pm ^{+0.04}_{-0.05}$ & $0.48 \pm ^{+0.02}_{-0.03}$ &$830$  & $770$ \\ 
All  & $0.48 \pm 0.04$ & $-$  & $-$  & $-$ \\ 
\hline
\hline
&& Early-Type (E+S0) &&\\
\hline
SN Ia & $0.32 \pm 0.05$ & $0.35 \pm ^{+0.05}_{-0.04}$ &$58$   & $40$ \\ 
SBF   & $0.28 \pm 0.04$ & $0.25 \pm ^{+0.03}_{-0.02}$ &$265$  & $241$ \\
SFI   & $0.30 \pm 0.04$ & $0.33 \pm ^{+0.02}_{-0.03}$ &$830$  & $784$ \\
All  & $0.30 \pm 0.03$ & $-$  & $-$  & $-$ \\
\hline

\end{tabular}

\label{tbl-2}
\end{center}
\end{table}

The results of the VELMOD and $\chi^2$ minimizations for the
morphological subsamples are given in Table \ref{tbl-2}. For the
VELMOD determinations there is excellent agreement between the values
of $\beta$ derived from each of the peculiar velocity data sets for
both early- and late-type density fields.  There is also good
agreement between the $\chi^2$ and VELMOD techniques. The early-type
density field yields $\beta$ values ($\sim 0.3$) that are consistently
lower than in the late-type case ($\sim 0.48$).  This suggests that
the early-type density field is more strongly clustered
$\beta\sbr{S}/\beta\sbr{E}=b\sbr{E}/b\sbr{S} \sim 1.6$ in the $K_s$-band.

In principle, it is possible to test whether early- or late-types are
better tracers of the mass by comparing the quality of the fit from
the two different density fields to the same peculiar velocity
sample. For the SN Ia and SBF data sets, the peculiar velocity field
predicted by early-types has a lower $\chi^2$ value, although this
difference is only significant for the SN Ia sample.  For the SFI data
set, the opposite is true: the late-type density field yields a better
goodness-of-fit.
One might be concerned that, because the early-type density field is
sparser (and hence more subject to shot noise), its predicted peculiar
velocities are noisier and hence the $\chi^2$ may be biased
high. However, the early-type galaxies are also more strongly
clustered, with the result is that the signal-to-noise ratios of the
density fields are similar. To put it another way, the higher degree
of noise in the density field is suppressed by the lower value of
$\beta$. We have performed tests to simulate the effects of sparse
sampling on the goodness-of-fit statistic $\chi^2$, and find no
conclusive evidence for a difference between early and late-type
density fields. However, further tests are required to understand the
systematics in greater detail.

\section{Discussion \label{Discussion}}

\subsection{Potential Systematic Effects}

We now discuss the potential systematics which might affect our
results. These include smoothing, external gravity contributions,
sparseness of the sample, as well as the possibility of inhomogeneous
Malmquist bias.

Berlind at al. (2001) explored the effects of applying various biasing
schemes and smoothing lengths in the determination of $\beta$. For the
case of linear biasing, they find that the optimal smoothing length is
roughly a $3 h^{-1}$Mpc Gaussian sphere, which is close to our adopted
$5$ $h^{-1}$Mpc top-hat smoothing. In contrast, for example, had we
adopted a top-hat smoothing of $8 h^{-1}$Mpc, the results of Berlind
et al. suggest that our derived value of $\beta$ would have been
biased too high by $\sim 10\%$. Therefore our choice of smoothing
length is close to optimal and we do not expect a significant bias in
our determination of $\beta$.

The external velocity field can be reasonably described by the two
main components (a dipole and a quadrupole, i.e. bulk and shear).
Since the analysis was carried out in the LG frame, we need not worry
about any external dipole contributions, however, this is not true for
any higher order contributions, such as the quadrupole, that may
exist.  Willick \& Strauss (1998) suggested the existence of a
residual quadrupole velocity field that was not well modelled by the
IRAS 1.2Jy gravity field.  We find no evidence of a residual
quadrupole in the SN Ia and SFI data.  The SBF data suggests a
marginally significant quadrupole, but the inclusion of such a term
does not significantly affect the value of $\beta$.

Another possible systematic effect is related to the discreteness of
the density fields used to make our velocity predictions. Since the
galaxy density field is a sparsely sampled, this can lead to shot
noise in the velocity predictions. To test whether the sample size
affects the determination of $\beta$, we generated a sparse-sampled
density field comprising of $50\%$ of the full 2MASS density
field. For this test, we find a small change ($\sim 2-3\%$) in the
predicted value of $\beta$, which is negligible compared to the random
errors on $\beta$.

Finally, let us discuss the corrections made for inhomogeneous
Malmquist bias (IMB), which is a Malmquist bias arising from distance
errors scattering peculiar velocity data away from overdensities.
Since their redshifts are not scattered significantly, the resulting
pattern mimics infall, leading to a biased estimate of $\beta$. The
strength of the bias increases as the square of the typical peculiar
velocity error, $\sigma_{d_c}$, thus we expect the SFI sample to be
most affected by the IMB. The VELMOD technique includes a
$(1+\delta\sbr{g})$ correction to the distance probability function $P(r)$
which corrects for this effect.  It is nevertheless of interest to
estimate the effect of the IMB on different samples, which can be done
by repeating the analysis omitting the $(1+\delta\sbr{g}$) term. For the SN
Ia and SBF samples, the correction is negligibly small (0.02) in
$\beta$. The SFI data set is more sensitive to IMB; our result for
this sample $\beta=0.55$ (with IMB correction) would have been
$\beta=0.71$ without the IMB correction. For deeper samples with large
errors, it is clearly important to account for IMB.

In summary, we have not identified any systematics which affect the
result at a level greater than the random errors.

\subsection{Comparison with Other Results}

This paper is the first comparison between the 2MASS density field and
peculiar velocity surveys. However, our results are consistent with
previous analyses comparing the 2MASS dipole to the motion of the
LG. Maller et al. (2003) assumed $\Omega\sbr{m} = 0.27$ and found $b_{K} =
1.06\pm0.17$ which is equivalent to $\beta_{K} = 0.48\pm0.08$.  Our
result is also consistent with the upper limit $\beta_{K} < 0.55\pm0.2$
derived by Erdogdu et al. (2005).

\begin{table*}[tp]
\begin{center}
\caption{A summary of the $\beta$ results from this paper and recent v-v determinations obtained from the literature}
\begin{tabular}{lccl  } 
\hline 
\hline
Comparison Sets & $\beta$ & $\left(\frac{\Omega\sbr{m}}{0.3}\right)^{0.6}\sigma_8$ & Reference \\ 
\hline
SN Ia $vs.$ 2MASS            & $0.47 \pm 0.06$ & $0.86 \pm 0.15$ & This Study  \\ 
SBF $vs.$ 2MASS             & $0.46 \pm 0.06$ & $0.84 \pm 0.15$ & This Study  \\ 
SFI $vs.$ 2MASS             & $0.55 \pm 0.05$ & $1.02 \pm 0.15$ & This Study  \\ 
\bf SN Ia+SBF+SFI $vs.$ 2MASS &$\bf 0.49 \pm 0.04$ &$\bf 0.91 \pm 0.12$ &\bf This Study  \\ 
\hline
Mark III $vs.$ IRAS 1.2 Jy  & $0.50 \pm 0.10$ & $0.82 \pm 0.16 $  & Davis et al. (1996)\\ 
SN Ia $vs.$ IRAS 1.2 Jy      & $0.40 \pm 0.15$ & $0.65 \pm 0.25 $   & Riess et al. (1997)   \\ 
SBF $vs.$ IRAS 1.2 Jy       & $0.42^{+0.10}_{-0.06}$& $0.70 \pm ^{+0.16}_{-0.10}$   & Blakeslee et al. (1999)   \\ 
SFI $vs.$ IRAS 1.2 Jy       & $0.60 \pm 0.10$ & $0.99 \pm 0.09$  & da Costa et al. (1998)\\
Mark III $vs.$ IRAS 1.2 Jy  & $0.50 \pm 0.04$ & $0.82 \pm 0.08$  & Willick \& Strauss (1998)  \\
Mark III  $vs.$ PSCz        & $0.60 \pm 0.10$ & $0.99 \pm 0.19$  & Saunders et al. (1999)  \\
ENEAR  $vs.$ PSCz           & $0.50 \pm 0.10$ & $0.82 \pm 0.16$  & Nusser et al. (2001)  \\
SN Ia $vs.$ PSCz             & $0.54 \pm 0.06$ & $0.89 \pm 0.10$   & Radburn-Smith et al. (2004)  \\
SFI  $vs.$ PSCz            & $0.42 \pm 0.04$ & $0.66 \pm 0.08$  & Branchini et al. (2001)  \\ 
SEcat  $vs.$ PSCz           & $0.51 \pm 0.06$ & $0.84 \pm 0.10$  & Zaroubi et al. (2002)  \\ 
\bf Weighted IRAS Avg.      & $\bf 0.50 \pm 0.02$ &$\bf 0.83 \pm 0.06$  &           - \\
\hline
\bf 2MASS and IRAS Avg.     & \mbox{} &$\bf 0.85 \pm 0.05$  &           - \\
\hline
\end{tabular}
\label{tbl-3}
\end{center}
\end{table*}

Table \ref{tbl-3} lists our results and recent $\beta$ measurements
obtained via velocity-velocity methods obtained using the IRAS 1.2 Jy
and PSCz surveys. The 2MASS $\beta$ values agree quite well with IRAS
values.  However, when comparing the 2MASS predictions of $\beta$ to
the IRAS values, one should keep in mind that the IRAS samples are
dominated by late-type galaxies.  Consequently, the strength of the
clustering, or bias, is likely to differ and the comparisons of
$\beta$ may not be straight forward. With this in mind, it is
preferable to transform the results into a form that is independent of
the details of the density field used.  This can be accomplished by
noting that the linear biasing factor $b$ relates the r.m.s. galaxy
fluctuations, $\sigma\sbr{8,g}$, to the amplitude of r.m.s. mass
fluctuations, $\sigma_8$, where the subscript 8 indicates that the
density fluctuations have been averaged in top-hat spheres of radius
$8$ $h^{-1}$Mpc. If $\sigma\sbr{8,g}$ is known, we may express our
results as $\beta\sbr{g}\sigma\sbr{8,g} =
\Omega\sbr{m}^{0.6}\sigma_8$.

From the 2MASS $K_s$ band density field reconstructed in this paper,
we have measured directly the r.m.s fluctuations in 8 $h^{-1}$ Mpc
spheres. Allowing for shot noise, we find a value of
$\sigma_{8,K}=0.90$ (in reconstructed ``real'' space).  We have not
measured the uncertainty in this quantity, but note that this result
is consistent with $\sigma_{8,K}=1.0 \pm 0.1$ found by Maller et
al. (2005) for a larger $K_s$ band 2MASS sample, although it is lower
than the value found by Frith et al. (2005): $1.16\pm0.04$.
Luminosity-dependent biasing may account for some of the difference
between these results. Here we adopt $\sigma_{8,K}=0.9\pm0.1$ to
obtain the result $\sigma_8(\Omega\sbr{m}/0.3)^{0.6} = 0.91\pm0.12$.
We note that the is dominant contribution to the uncertainty is from
the uncertainty in $\sigma_{8,K}$.  We expect that $\sigma_{8,K}$ will
be better determined in the near future from the 2MASS Redshift Survey
(Huchra 2000).

For the IRAS survey, Hamilton \& Tegmark (2002) found that
$\sigma_{8,I}=0.80 \pm 0.05$. We quote
$\sigma_8(\Omega\sbr{m}/0.3)^{0.6}$ values in the third column of
Table \ref{tbl-3}. For the IRAS case, the v-v results have been
combined to yield an average IRAS value
$\sigma_8(\Omega\sbr{m}/0.3)^{0.6} = 0.83 \pm 0.06$.  There is good
agreement between the various methods and data sets used in $\beta$
determinations from peculiar velocity studies.  The IRAS result is
tighter than our result because $\sigma_{8,I}$ is better constrained
and more peculiar velocity data have been compared to IRAS
predictions. A weighted average of IRAS and 2MASS results gives
$\sigma_8(\Omega\sbr{m}/0.3)^{0.6} = 0.85\pm0.05$.

Feldman et al. (2003) measured the mean relative peculiar velocity for
pairs of galaxies, and in comparison with the correlation function,
derived a value of $\sigma_8=1.13 \pm 0.22$ for
$\Omega\sbr{m}=0.30^{+0.17}_{-0.07}$, which is consistent with our
result. Feldman et al. do not quote a result for the combination
$\Omega^{0.6} \sigma_8$, which one would expect to have smaller
errors, than the marginal errors on $\sigma_8$ and $\Omega\sbr{m}$
separately.

Unlike the v-v comparisons, values can also be obtained by
density-density ($\delta-\delta$) comparisons. Previously, this method
required the construction of the gravitation potential from the radial
velocity field (POTENT; Bertschinger \& Dekel 1989), differentiation
and comparison it to the galaxy density field derived from redshift
survey data. Note that such comparisons are based on smoothing and
differentiating sparse and noisy peculiar velocity data. Using this
method, Sigad et al. (1998) found $\beta\sbr{I}=0.89\pm0.12$.  A newer
method based on an unbiased variant of the Weiner filter yields values
that are more consistent with those of the v-v method: $\beta\sbr{I}
=0.57\pm0.12$ (Zaroubi et al. 2002).

It is possible to use peculiar velocities to obtain estimates of
$\beta$ in a completely different way.  Peculiar velocities of
galaxies distort the pattern of galaxy clustering in redshift space,
making the redshift space power spectrum anisotropic. One can use the
distortions to measure the parameter $\beta$, and with information
about the density field obtain the cosmological constraint
$\sigma_8\Omega\sbr{m}^{0.6}$.  We combine the measured
$\sigma\sbr{8,g}$ for the 2-degree Field Galaxy redshift Survey
(2dFGRS) by Lahav et al. (2002) with the 2dFGRS $\beta$ result of
Hawkins et al. (2003) to obtain the combination
$\sigma_8(\Omega\sbr{m}/0.3)^{0.6}=0.89 \pm 0.23$.  Similarly,
Percival et al. (2004) used a spherical harmonics method on the 2dFGRS
to find $\sigma_8(\Omega\sbr{m}/0.3)^{0.6}=0.95 \pm 0.12$. These are
in good agreement with our result.  

The comparisons between peculiar velocity derived values of
$\sigma_8\Omega\sbr{m}^{0.6}$ can be extended to other cosmological
data. For instance, Refregier (2003) compiled an average
$\sigma_8=0.83 \pm 0.04$ value based on many weak lensing survey
results for the assumed cosmology defined by $\Omega\sbr{m}=0.3$,
$\Omega_{\Lambda}=0.7$, $\Gamma=0.21$.  Contaldi et al. (2003) used
Cosmic Microwave Background (CMB) plus weak lensing results to
constrain the combination at $\sigma_8(\Omega\sbr{m}/0.3)^{0.5}=0.88
\pm 0.04$.  Tegmark et al. (2004) used recent CMB and Sloan Digital
Sky Survey (SDSS) power spectra to find
$\sigma_8(\Omega\sbr{m}/0.3)^{0.6}=0.95\pm 0.13$ under the assumption
of a flat, $n=1$ universe. Similarly, Seljak et al. (2005), combined
WMAP with SDSS clustering, bias and Ly$\alpha$ forest data to obtain
$\sigma_8(\Omega\sbr{m}/0.3)^{0.6}=0.86\pm0.05$ (assuming a flat
universe with free $n$).  Using CMB and 2dFGRS data, Sanchez et
al. (2005) find $\sigma_8(\Omega\sbr{m}/0.3)^{0.6}= 0.71\pm0.10$
assuming (flat, $n=1$).  Sanchez et al.\ do not quote errors for the
combination $\Omega^{0.6}\sigma_8$, so above we have calculated the
error assuming that the uncertainty in $\Omega\sbr{m}$ and $\sigma_8$
are independent.  Since their Figure 1 indicates that these parameters
are not independent, the above uncertainty will be an overestimate.
Although the conflict is not significant, it is interesting that the
Sanchez et al. result is lower than the CMB plus SDSS values, the
2dFGRS redshift-space distortion results, as well as our result.

Overall, our peculiar velocity results are in good agreement with a
broad variety of recent $\sigma_8 \Omega\sbr{m}^{0.6}$ measurements
from the literature.

\subsection{Implications for Large-Scale Flows}

Although the focus of this paper has been on the gravity and peculiar
velocity field within $R\sbr{max}$, there are important gravitational
sources beyond this limit. While the existence of these sources do not
affect our $\beta_{K}$ results (as we have preformed the analysis in
the LG frame where external contributions are negligible), we can turn
the problem around and place constraints on the large scale flow using
our value of $\beta_{K}$.

Specifically, for the best fitting $\beta_{K}=0.49\pm0.04$, we
calculate a LG velocity $\mathbf{v}(0)=403 \pm 27$ km s$^{-1}$ in the
direction $l=257^{\circ}$, $b=37^{\circ}$. The uncertainty on this
linear theory predict ion is underestimated, because we expect that
part of the LG's motion is not well described by linear theory.  Since
the LG is a galaxy group, and the relative infall of the Milky Way and
Andromeda has already been accounted for following Yahil et
al. (1977), it should be less affected by nonlinearities than are
individual galaxies. It seems reasonable to adopt a thermal velocity
dispersion of $100$ km s$^{-1}$, making $v(o)$ uncertain at the $25\%$
level. Thus for our best fit $\beta$, the local volume does not fully
account for the LG's motion with respect to the CMB, and there is a
residual velocity dipole of $\mathbf{U}=271 \pm 104$ km s$^{-1}$ in
the direction $l=300^{\circ}$, $b=15^{\circ}$.

This residual dipole is slightly lower than, but consistent with, the
result of Hudson (1994b), who compared the predictions of an
optically-selected galaxy density field to $D_n-\sigma$ and
Tully-Fisher peculiar velocities and found $\beta\sbr{o}=0.50 \pm
0.06$ and a residual velocity, from beyond $80h^{-1}$Mpc, of $405 \pm
45$ km s$^{-1}$ toward $l=292^{\circ}$, $b=7^{\circ}$. The residual LG
dipole found here is in good agreement with a recent estimate of the
bulk flow of $\mathbf{U}=225$ km s$^{-1}$ toward $l=300^{\circ}$,
$b=10^{\circ}$ found by Hudson et al. (2004), for a sample of peculiar
velocities at depths greater than $60 h^{-1}$ Mpc.

Thus, our derived $\beta_{K}$ suggests that there are significant
contributions ($\sim 40\pm15$\%) to the LG's motion arising from
sources at $r > 65h^{-1}$Mpc.

\section{Conclusions}

In this paper we have used the 2MASS catalog and public redshift data
to reconstruct the local density field. Under the assumption of GI and
we used linear theory to derive the peculiar velocity fields which we
compared to measured peculiar velocity data.  The different data sets
all yield results from the VELMOD method that are in very good
agreement, with a best fit $\beta=0.49 \pm 0.04$.  Calculation of the
r.m.s. density fluctuations of K-selected galaxies allowed us to
generalize the result as $\sigma_8(\Omega\sbr{m}/0.3)^{0.6}=0.91\pm
0.12$. Our result is consistent with previous determinations of
redshift space distortion methods, CMB and weak lensing results. The
low value of $\beta$ suggests that significant contributions to the
LG's motion arise from beyond $65h^{-1}$Mpc.

We would like to stress the power of the method that we have outlined
here. Peculiar velocity comparisons provide direct and independent
measures of $\sigma_8\Omega\sbr{m}^{0.6}$, which are consistent with
results from a wide range of techniques. Peculiar velocities probe
scales near the fiducial $8 h^{-1}$Mpc scale and allows us to make
these predictions without any assumptions regarding the background
cosmology. All that is required is a source of redshift data, and a
sample of galaxies which have redshift-independent distance
information. This illustrates the power of peculiar velocity studies
to probe the underlying cosmology.

In future peculiar velocity work, it may be possible to break the
degeneracy between $\Omega\sbr{m}^{0.6}$ and $b$ by applying a more
sophisticated biasing model, such as a nonlinear bias based on the
{\it halo model} (e.g.\ Marinoni \& Hudson 2002).  Improving the
predictions in the mildly nonlinear regime (e.g.\ Nusser \& Branchini
2000) will be important for this latter goal.

Furthermore, a deeper and more complete 2MASS data set will reduce the
errors associated with sparseness and enable a better reconstruction
of the local density field.  Similar improvement will come with more
reliable distance estimates and larger peculiar velocity surveys.  For
instance, Wang et al. (2005) are currently working on compiling a
sample of SN Ia that have distance errors, as low as $3-4\%$, although
the sample is still quite small at present.  As the data and
techniques continue to improve, the future of peculiar velocity
studies will continue to be a promising area for understanding the
mass distribution in the local universe.

\section{Acknowledgements}

The authors would like to thank Riccardo Giovanelli for providing us
with a tabulated version of the SFI peculiar velocities.  We also
thank Christian Marinoni for interesting discussions in the early
phases of this project.  MJH acknowledges support from the NSERC of
Canada and a Premier's Research Excellence Award.

\clearpage

\end{document}